\documentclass[useAMS,usenatbib,usegraphicx,iop,appendixfloats]{emulateapj}

\usepackage{verbatim}
\usepackage{xspace}
\usepackage{rotating}
\usepackage{array}
\usepackage{url}

\def\halpha{H$\alpha$\xspace}

\def\micron{$\mu$m\xspace}
\def\hii{H$\,$\textsc{ii}\xspace}
\def\hi{H$\,$\textsc{i}\xspace}

\def\Msun{M$_{\sun}$\xspace}
\def\msun{M$_{\sun}$\xspace}

\def\flam{F$_{\mathrm{\lambda}}$\xspace}

\shorttitle{Interarm FUV emission in M101}

\begin{document}

\title{Origin of the Diffuse, Far Ultraviolet Emission in the Interarm Regions of M101 }

\author{Alison F. Crocker\altaffilmark{1\dag, 2},
Rupali Chandar\altaffilmark{2},
Daniela Calzetti\altaffilmark{3},
Benne Willem Holwerda\altaffilmark{4},
Claus Leitherer\altaffilmark{5},
Cristina Popescu\altaffilmark{6},
R. J. Tuffs\altaffilmark{7}}

\altaffiltext{1}{Department of Physics, Reed College, Portland, OR, 97202}
\altaffiltext{2}{Department of Physics and Astronomy, University of Toledo, Toledo, OH, 43606}
\altaffiltext{3}{Department of Astrophysics, University of Massachusetts, 710 North Pleasant Street, Amherst, MA, 01003}
\altaffiltext{4}{University of Leiden, Leiden Observatory, Niels Bohrweg 4, NL-2333, Leiden, The Netherlands}
\altaffiltext{5}{Space Telescope Science Institute, 3700 San Martin Drive, Baltimore, MD 21218, USA}
\altaffiltext{6}{Jeremiah Horrocks Institute, University of Central Lancashire, Preston PR1 2HE, UK}
\altaffiltext{7}{Max Planck Institut fŸr Kernphysik, Saupfercheckweg 1, D-69117 Heidelberg, Germany}
\altaffiltext{\dag}{crockera@reed.edu}

\begin{abstract}   

We present images from the Solar Blind Channel on {\em HST} that resolve hundreds of far ultraviolet (FUV) emitting stars in two  $\sim$1 kpc$^{2}$ interarm regions of the grand-design spiral M101. 
The luminosity functions of these stars
are compared with predicted distributions from simple star formation histories,
and are best reproduced when the star formation rate has 
declined recently (past $10-50$~Myr).
This pattern is consistent with stars forming within spiral arms and
then streaming into the interarm regions. We measure the diffuse FUV surface brightness after subtracting all of the detected stars, clusters and background galaxies. A residual flux is found for both regions which can be explained by a mix of stars below our detection limit and scattered FUV light. The amount of scattered light required is much larger for the region immediately adjacent to a spiral arm, a bright source of FUV photons.

\end{abstract}

\section{Introduction}

Numerous massive O and B-type stars, typically in stellar clusters and associations, form in the spiral arms of spiral galaxies. 
These stars emit strongly in the UV and thus strikingly delineate these spiral arms or armlets. However, UV emission is also present in the interarm regions of spirals \citep[e.g.,][]{marcum01}, which is commonly called ``diffuse'' emission because of its smooth appearance in Ultraviolet Imaging Telescope (UIT) or Galaxy Evolution Explorer (GALEX) images. Three distinct phenomena likely contribute to this interarm UV flux. First, stars formed within the arms should disperse into the interarm regions due to their streaming motion with respect to the arms. Because O stars only live for $\sim10$~Myr, these dispersed stars should be predominantly B-type stars. Second, low-level star-formation may occur {\it in situ} within the interarm regions, directly contributing young UV-bright stars (the initial mass function [IMF] may or may not be the same as within the spiral arms). Third, dust within the interarm regions should scatter UV light from the strongly star-forming arms into the observed line-of-sight. 

UIT images originally showed that the FUV-NUV color of interarm regions is redder than that of the spiral arms \citep{landsman92, cornett94}. The simplest explanation for this phenomenon is that the luminosity-weighted age of the stellar populations is older in the interarm regions than within the arms, as expected if spiral arms are the locations of increased star formation.

However, scattered light is also expected to contribute to the interarm UV flux. Monte Carlo-based radiative transfer models show that scattered flux in the FUV can range from approximately 0.1 to 1 times the measured stellar flux, depending on the type of dust, the dust geometry, the dust clumpiness, and the optical depth to the midplane of stellar sources \citep{witt00}. Additionally, a higher fraction of scattered light is expected when the viewing direction is close to perpendicular to the stellar disk, as is typically the case for measuring interarm light.

These model predictions fit with observations of the SMC and LMC which show that detected stars only make up 55\% and 75\% of the total FUV light, respectively \citep{cornett97, parker98}. Detection of linearly polarized light from the Wide-Field Imaging Survey Polarimeter (WISP) demonstrates that a significant portion of this unaccounted-for light is scattered (although exactly how much depends on dust-source relative distribution and the scattering phase function; \citealp{cole99a}). 
 
For UV photons to be scattered effectively, the optical depth must be relatively low. As the optical depth increases, multiple dust interactions lead to a higher fraction of light absorbed. A low UV optical depth requires a low dust surface density, most likely found in the interarm or outer parts of disk galaxies \citep[e.g.,][]{holwerda05} or in their halos, extending away from the disk midplane \citep{hodges-kluck14,seon14}.   \citet{thilker05a} found that the proportion of diffuse UV emission increased at large radii in the flocculent spiral M33, as expected for scattered light. Additionally, diffuse UV emission is seen in the halos of the starbursts NGC~253 and M82, far from their stellar disks \citep{marcum01, hoopes05}. This emission is most likely dust-scattered UV from their starburst cores. 

Comparing GALEX images with data from the Infrared Space Observatory (ISO), \citet{popescu05} found that the far-infrared (FIR) to UV ratio is higher in the interarm regions than within spiral arms in M101 and argue that this is best explained by dust-scattered light in the FUV. However, other explanations for a boosted interarm FIR to UV ratio are possible, including a shift to a slightly older stellar population. Here, we directly investigate the scattering hypothesis to explain the interarm FUV light in M101. Our method relies on the excellent resolution of the Hubble Space Telescope's (HST) Solar Blind Channel (SBC) detector, which resolves individual UV-bright stars at the 6.7~Mpc distance of M101. 

In Section~2, we describe the SBC data and our photometric measurements of point-like sources from both the SBC data and optical Wide Field Camera (WFC) data. 
Section~3 presents our main analysis of the FUV-emitting stellar populations, including the recent star formation histories (SFHs) of our observed regions, color-magnitude diagrams and identification of stellar clusters. In Section~4, we evaluate the sources of the significant amount of remaining FUV emission in the two fields. Section~5 presents our conclusions.

\section{Observations and data reduction}

\subsection{FUV Observations}

\begin{figure}
\begin{center}
\includegraphics[width=9.2cm]{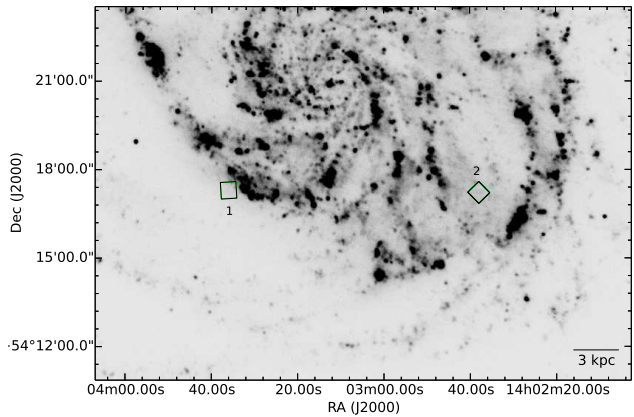}
\caption{The two SBC pointings (small squares) over the GALEX FUV image of M101. 
}
\label{fig:Uband}
\end{center}
\end{figure}

Using the SBC detector of the Advanced Camera for Surveys \citep{ford98} on the $HST$,
we obtained deep, far-UV images with a $34.6\arcsec \times 30.8\arcsec$ field-of-view 
of two interarm regions within M101.
These two pointings are shown in Fig.~\ref{fig:Uband}.
The images were taken with the F150LP long-pass filter, which has  
an effective wavelength of 1614\AA\xspace and a FWHM of 177\AA.
Pointing~1 was observed for a total of 10,200~s in a single visit of 16 exposures.
Pointing~2 was observed for a total of 10,950~s in two visits of 9 and 8 exposures each.
Small dithers were applied between individual exposures in order to reduce the
impact of hot and warm pixels.

\begin{figure*}
\begin{center}
\includegraphics[width=8.95cm]{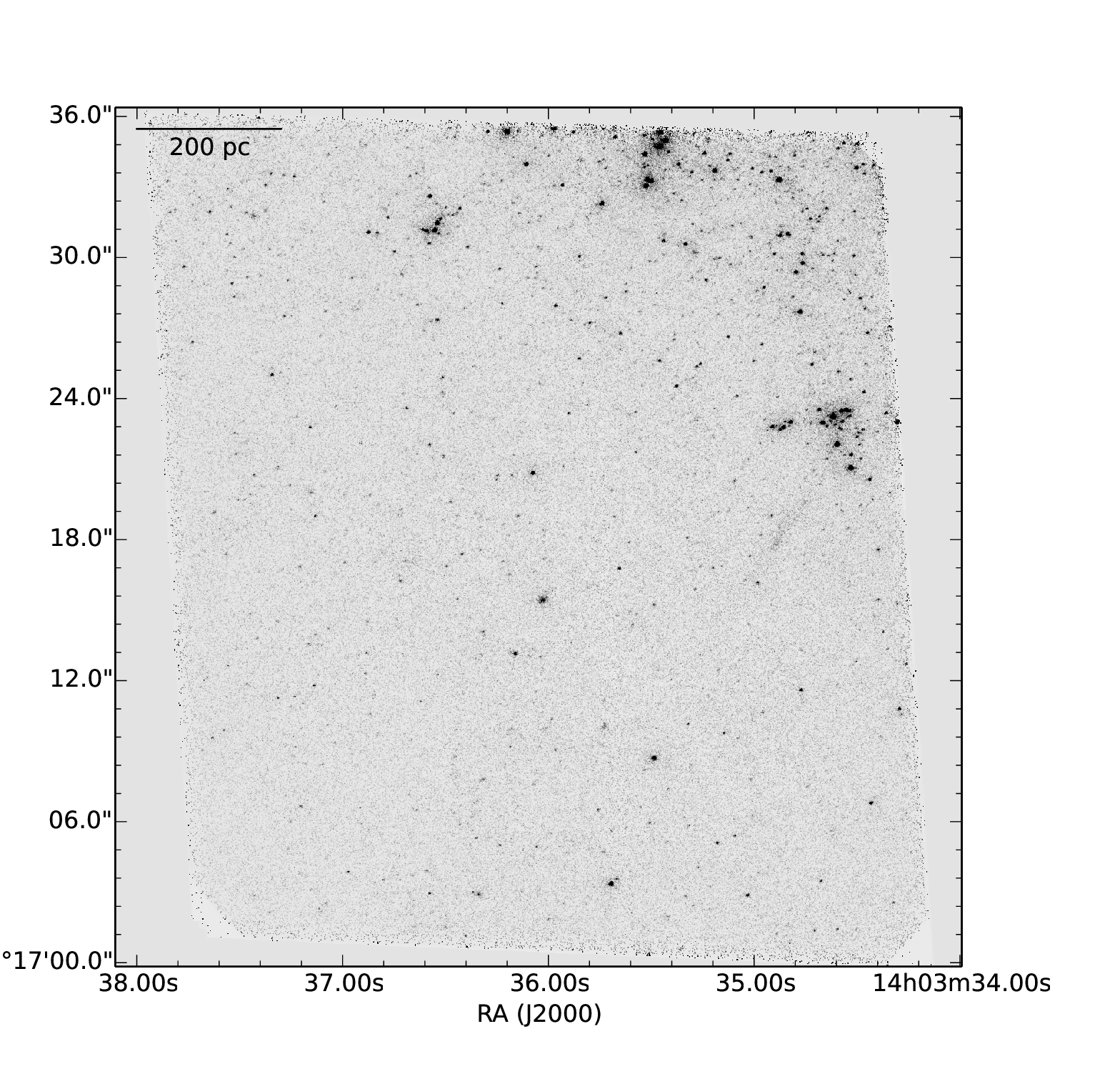}
\includegraphics[width=8.95cm]{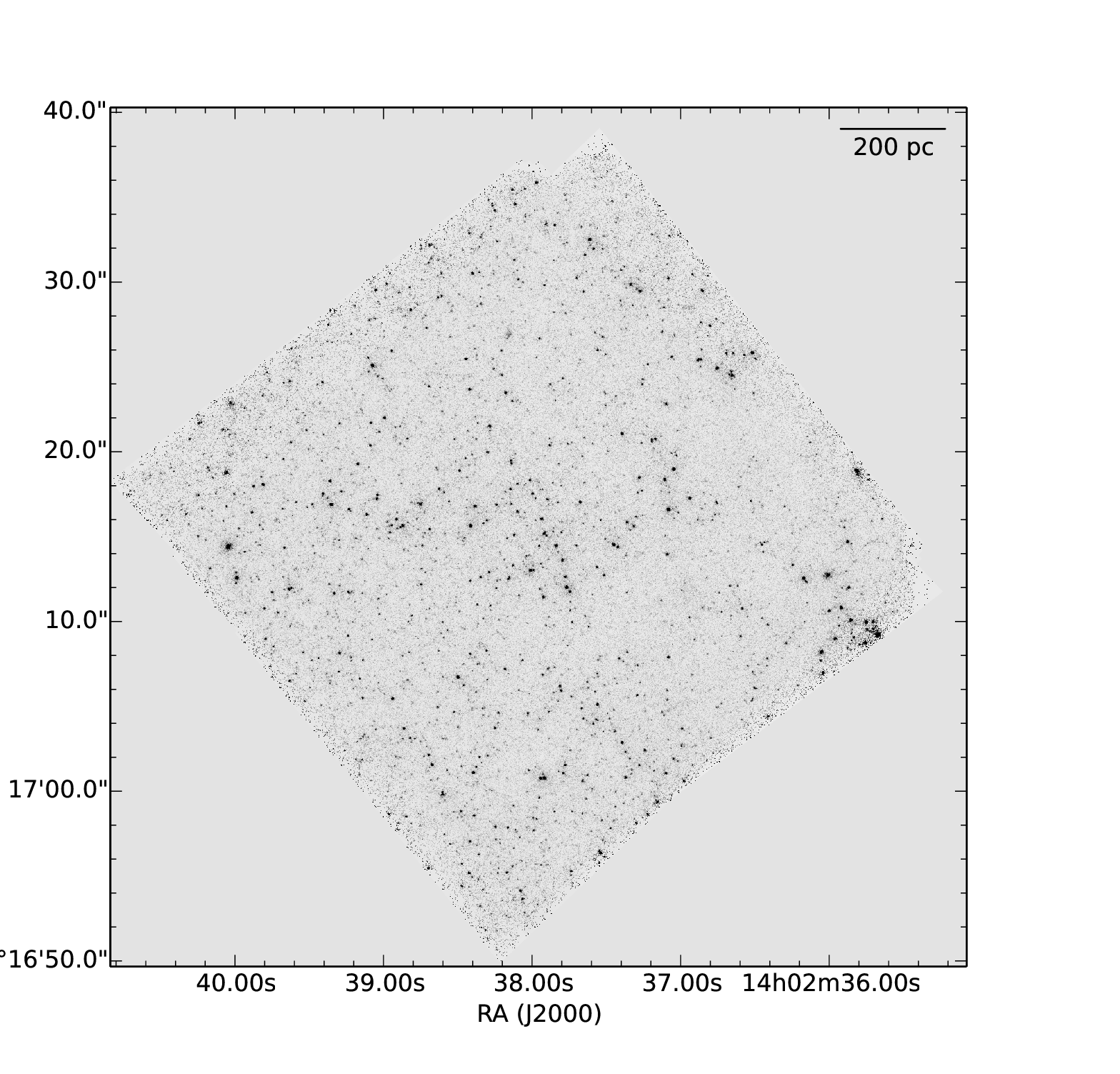}

\caption{The SBC F150LP images of Pointing 1 (left) and Pointing 2 (right).
}
\label{fig:SBCimages}
\end{center}
\end{figure*}

While there is no read noise in the SBC detector, it does have a dark current that must be
subtracted before measuring the diffuse light. The detailed procedure we followed to remove the dark current from our data is described in Appendix~A and summarized here. We find that the dark current can be modeled by two components, one that is always present (``low temperature") and a second that appears only at higher detector temperatures (significant for approximately half of our data). 
Both components contain structure. The high temperature structure peaks towards the center of the detector
(see also \citet{teplitz06}), while the low temperature structure has a more modest
gradient, increasing from the left to right hand side of the detector. We first create maps of the high temperature dark current, using 
the SBC observations themselves, by subtracting an average and spatially smoothed low-temperature image from a similar high temperature image, leaving a map of only the high temperature component.
This high temperature component is then removed from all appropriate individual frames, leaving only the cold temperature component. All frames are then combined together. The low temperature dark current is determined by comparing resolution matched SBC images to those of the same location taken with GALEX in the far-UV.

After we remove the dark current, we combine all of the individual
exposures together to make a single image of each field using
DrizzlePac's ASTRODRIZZLE with 0.025\arcsec\xspace pixels. 
A WCS solution used to align the two different sets of observations 
for Pointing~2 was computed using DrizzlePac's TWEAKREG (STSci PyRAF package).
The resulting far-UV images are shown in Fig.~\ref{fig:SBCimages}.

\subsection{FUV photometry}

We identify point and point-like sources in the combined far-UV images
using the DAOFIND task in IRAF.
We use a detection limit of 4.5 counts, since the background level is dominated by
Poisson noise rather than read noise (observed to be below 0.5 counts per pixel). At the level of 4.5 counts, 
we are almost certainly detecting real sources ($P < 0.02\%$).
We found that we needed to run DAOFIND with two different sets of parameters, in order to
detect all obvious sources. % seen in the far-UV images, particularly fainter sources.
%In order to recover both point and somewhat extended sources, we run DAOFIND twice. 
In the first run, the FWHM of the convolution kernel is set to 1.5 times that of the PSF (as commonly recommended for point-source detection in the read-noise limited case) and in the second to 2.5 times that of the PSF. 
Combining the two DAOFIND runs for each pointing results in initial source catalogs that include 709 objects in Pointing 1 and 1523 objects in Pointing 2.

\begin{figure}
\begin{center}
\includegraphics[width=8cm]{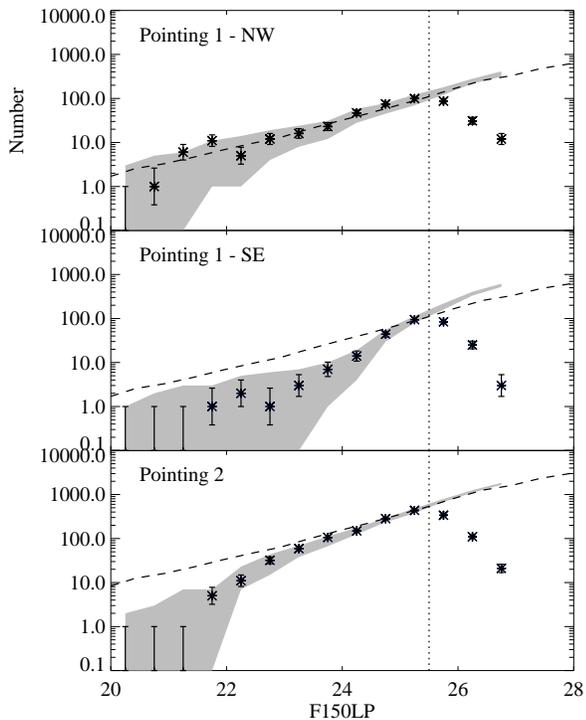}
\caption{Distributions of F150LP apparent magnitudes of detected stellar sources (black points). The dashed black line represents the expected F150LP magnitude distribution for a constant SFR (values given in Table~1). The grey band represents the range of random samples pulled from the best-fit distribution of SFH for each region. The vertical dotted line shows our estimated completeness limit at 25.5 mag.
}
\label{fig:lumfunc}
\end{center}
\end{figure}

We perform aperture photometry with DAOFIND's PHOT task using a 4 pixel radius aperture,
and use a zeropoint of 20.4423~mag from the ACS Zeropoints website\footnote{http://www.stsci.edu/hst/acs/analysis/zeropoints/\#sbc}, in order to convert to the VEGAMAG  photometric system. 
We apply an aperture correction of $-0.651$~mag, determined from photometry of bright,
isolated stars in the SBC/F150LP observations of globular cluster NGC~6681 (Proposal ID = 11378).
The observed luminosity function of the point (stellar) sources are shown as the data points in Fig.~\ref{fig:lumfunc}. 
The distributions increase steadily down to $\sim25.5$~mag; at fainter magnitudes than this, the source lists become incomplete.

\subsection{Optical Observations and Photometry}

Optical images of M101, covering the entirety of Pointings 1 and 2,
were taken with the Wide Field Camera on the ACS
as part of Proposal ID 9490 (PI: Kunz). 
We obtained drizzled images in the following filters: F435W ($\sim B$), 
F555W ($\sim V$), and F814W ($\sim I$) from the Hubble Legacy Archive.
The F435W and F150LP images were aligned by identifying 20 bright
stars in common, and using the IRAF task CCMAP to determine the offset
between the optical and far-UV images.
We perform photometry on the optical images using a 3 pixel radius aperture for all sources brighter than $m_{F150W} = 25$.
Zeropoints of 25.791 (F435W), 25.738 (F555W), and 25.533 (F814W) are 
used, as recommended by the ACS Zeropoint Calculator website\footnote{http://www.stsci.edu/hst/acs/analysis/zeropoints/zpt.py}.
Aperture corrections of $-0.28$, $-0.25$, and $-0.29$ were applied to the F435W,
F555W, and F814W filters respectively, based on the encircled energy distributions
presented in the ACS Instrument Handbook \citep{ubeda12}.
We only include optical photometry if the uncertainties are less than 0.2~mag.
Almost all sources brighter than $m_{F150LP} = 23$ have good photometric
measurements in all three optical bands, 
while sources with $m_{F150LP} \sim 25$ only have good measurements
35\%, 25\% and 20\% of the time in the F435W, F555W, and F814W images, respectively.

\section{Analysis of FUV Emitting Sources}

In this section, we analyze the various sources of far-UV emission in our interarm M101 fields: 
individual stars, stellar clusters, and diffusely distributed light.

\subsection{Stars}

All F150LP sources without an optical counterpart in F555W are considered to be stellar sources because any star cluster or galaxy contaminants would be detected in F555W. Specifically, the F555W images are sensitive to clusters down to 10$^{3}$ \Msun and background galaxies will have F150LP-F555W $> 0$ and thus be detected given the completeness limits. 

For F150LP sources with optical counterparts, we separate stars from stellar clusters and background galaxies based on their size, since all but the most compact clusters are expected to be broader than the point spread function at the distance of M101 (physical FWHM of 3.2~pc). Source sizes were measured in the F555W image, because the signal-to-noise is generally higher in this filter than in the FUV. Furthermore, mass segregation may cause the most massive stars (which emit strongly in the FUV) to sink to the centers of clusters, thereby  leading compact clusters that have narrower FUV than optical sizes.
In F555W, point sources should have a FWHM close to 2.0 pixels.
Based on measurements of a hand-selected sample of stars,
we find that point sources have measured FWHM $ < 2.35$~pix. We adopt this simple FWHM approach because standard star-identification parameters (DAOPHOT's ``sharp'' and ``round'') do not work well in this extremely count-limited case. Furthermore, \citet{holwerda14} demonstrate that source classification based on half-light radius (directly linked to FWHM for point sources) is more effective than other point-source identification techniques. 
%We also visually inspected each FUV source in all four available filters, and reject any that are obviously extended.
Thus our source catalogs are dominated by individual stars rather than by stellar clusters.

The shape of the stellar luminosity function in the FUV
provides important constraints on the recent star formation history.
The observed luminosity functions of the stellar sources are shown 
as the data points in Fig.~\ref{fig:lumfunc}, 
with Poisson uncertainties.
We have separated Pointing~1 into two distinct regions, since the north-west corner
of this field contains the edge of a spiral arm with a markedly different luminosity function.
The luminosity functions all increase down to the approximate 
completeness limit of 25.5~mag (shown as the dashed line in Fig.~\ref{fig:lumfunc}).

The shape of the stellar FUV luminosity function is driven by the recent star formation history. In order to interpret our observed distributions, we create predicted luminosity functions that assume different star-formation histories. We populate these theoretical luminosity functions star-by-star, first drawing the mass of the star from a Kroupa IMF, then assigning the age of the star according to the assumed star-formation history. We parametrize this recent star-formation history with a simple two-phase 
model with three parameters: the initial SFR, the final SFR, and a break time at which the 
SFR instantaneously changes from its initial to final value. This form of the SFH also provides an appropriate description for a scenario where young stars stream out of a star-forming arm into the inter-arm regions we have targeted. (In this case, the initial SFR maps to the within-arm SFR, the final SFR to a continuous level of SFR that may persist in the inter arm region itself and the break time to the time  since the average FUV-bright star passed out of the spiral arm.) 

We use Geneva stellar evolutionary models to assign a F150LP magnitude to each star given its mass and age. 
The model isochrones provide log surface gravity ($g$), effective temperature ($T_{\mathrm {eff}}$) and absolute V-band magnitude ($M_{\mathrm V}$)  values for stars between 0.8 and 120 \Msun at various ages\footnote{We use the half-solar metallicity (z=0.008, \citealp{lejeune01}) models as appropriate based upon M101's metallicity gradient \citep[e.g.,][]{li_y13}.}. Synthesized spectra from Kurucz stellar atmosphere models \citep[ATLAS9,][]{kurucz93} then help us bridge the gap to a F150LP magnitude. As we are not concerned with absorption lines nor the ionizing continuum, the Kurucz models should give nearly identical results to those of a non-LTE model such as TLUSTY for the B and early O stars ($< 30$ \msun) of our sample (for comparisons and discussion, see \citet{lanz07} and \citet{przybilla11}). Given this equivalence, the Kurucz models were used for simplicity as they are included in the STSDAS synphot package.
 %{\bf While not as accurate as a non-LTE model for lines and the ionizing continuum, there is a very good match between the non-ionizing continuum between the LTE Kurucz models and the non-LTE TLUSTY models for B stars as shown in Fig. 6 of \citet{lanz07}. We note that our sample does not include very massive O stars ($> 30$ \msun) for which the Kurucz models do not match the non-LTE TLUSTY models in the FUV.  Thus we consider the Kurucz models sufficient for our purpose and}
 We select the appropriate stellar atmosphere spectrum for each star based upon its assigned $\log(g)$ and $T_{\mathrm {eff}}$ and then compute the F150LP - V color by convolving this spectrum with Johnson V and F150LP filters using synphot. With this color, we convert the $M_{\mathrm V}$ from the model isochrone into a F150LP magnitude.
Binning these magnitudes, we obtain theoretical luminosity functions.

We parametrize our recent star-formation histories with different break times ($t_{\mathrm{break}}$) of  6, 9, 12, 16, 20, 25, 30, 40, 50, 60, 70, 85~Myr and ratios $\mathrm{SFR}_{\mathrm{initial}}$/$\mathrm{SFR}_{\mathrm{final}}$ of 0.01, 0.032, 0.1, 0.32, 3.2, 10, 32 and 100. Two additional cases are considered: (1) A truncated recent SFH, with SFH$_{\mathrm{final}}=0$ (which corresponds to no in-situ star formation in the inter-arm regions), and (2) a constant rate of recent star formation (which corresponds to a case where the SFR does not increase due to passage through an arm). 
We normalize $\mathrm{SFR}_{\mathrm{initial}}$ such that the number of predicted and observed sources match in the 25-25.5 magnitude bin. 
Table~\ref{tab:bestfit} lists the best-fit parameters for each location and each type of recent star formation history (broken, truncated, constant).  The best fit was determined by minimizing the Poisson fit statistic: 
\begin{equation}
-2\; \mathrm{ln} \;P = 2 \sum m_{i} - n_{i} + n_{i}\;\mathrm{ln}\frac{n_{i}}{m_{i}},
\end{equation}
from \citet{dolphin02b}. In this equation, we sum over each magnitude bin, where $n_{i}$ and $m_{i}$ represent the observed and predicted counts per bin, respectively. We fit using the magnitude bins up to 25.5. The grey bands in Fig.~\ref{fig:lumfunc} represent the spread in fifty magnitude distributions randomly sampled from the best-fit parameters.

 \begin{table}[htdp]
\begin{center}
\caption{Star formation history best fits}
\begin{tabular}{lcccc}
\hline
Fit Type & SFR$_{\mathrm{initial}}$ & SFR$_{\mathrm{final}}$ & $t_{\mathrm{break}}$ & $-2\mathrm{ln} P$\\ 
		& (\Msun yr$^{-1}$) & (\Msun yr$^{-1}$) & (Myr)\\
\hline % multiplicative amount of 24-25 flux to account for sub-det (fainter than 25) flux
\em{Pointing 1 - NW} \\
\phantom{000}Broken & 0.0073 & 0.00073 & 85 & 4.8\\ %???
\phantom{000}Truncated & 0.00076 & 0 & 6 & 9.7\\ % 4.28
\phantom{000}Constant & 0.00073 & --  & -- & 5.6\\ % 4.03
\em{Pointing 1 - SE} \\
\phantom{000}Broken & 0.00157 & 0.00016 & 40 & 0.5\\ % 16.28
\phantom{000}Truncated & 0.00084 & 0 & 12 & 24.0 \\ % 4.82
\phantom{000}Constant  & 0.00069 & -- & -- & 42.4\\ %4.03
\em{Pointing 2} \\
\phantom{000}Broken & 0.0037 & 0.00037 & 12 & 3.5\\ %4.78
\phantom{000}Truncated & 0.0036 & 0 & 9 & 7.0 \\ % 4.55
\phantom{000}Constant  & 0.0032 & -- & -- & 36.0\\  % 4.03
\hline
\label{tab:bestfit}
\end{tabular}
\end{center}
\end{table}%

 Our star-formation history fits indicate that the two interarm regions (Pointing 1 - SE and Pointing 2) have very low amounts of ongoing star-formation consistent with SFRs being lower by a factor $\approx 10$ in the inter-arm regions when compared to those within an arm. Pointing 2 is best fit by a star-formation history that breaks at 12~Myr, decreasing by tenfold at that time. Pointing 1 - SE requires a an older break time of 40~Myr, but also a very low current SFR. The region corresponding to the edge of a spiral arm (Pointing 1 - NW) has  the largest current SFR (despite covering the smallest area) and is well fit by a star-formation history that is either constant or has been constant for about the past 85~Myr. Fitting with a Salpeter IMF instead of a Kroupa IMF does not change the best-fit SFHs, as expected because these two IMFs only differ below 0.5 \Msun where stars do not emit appreciably in the FUV. We note that SFRs would be approximately 1.6 times higher if we had assumed a Salpeter IMF rather than a Kroupa IMF.
 
Faint dust emission from Spitzer and Herschel data \citep{dale07, gordon08, dale12} prompts us to consider the effect of extinction on our UV luminosity functions. Dust surface densities in the two regions range from 0.5 to $2\times 10^{5}$ \Msun kpc$^{-2}$ (Aniano et al., in prep.), corresponding to an $A_{\mathrm V}$ = 0.1-0.4 by the empirical conversion of \citet{kreckel13}. This conversion is valid for \hii regions, so based upon \citet{calzetti94}, we expect a lower average extinction towards the stars ($A_{\mathrm V}$ = 0.06-0.26). We assume a typical value of $A_{\mathrm V}$ = 0.1 and convert this to $A_{\mathrm{F150LP}}$ = 0.26 using the \citet{cardelli89} curve with an $R_{\mathrm V} = 3.1$. Applying this extinction to the theoretical models and refitting, the best fits for all three regions remain the same.

\begin{figure*}
\begin{center}

\includegraphics[width=8cm]{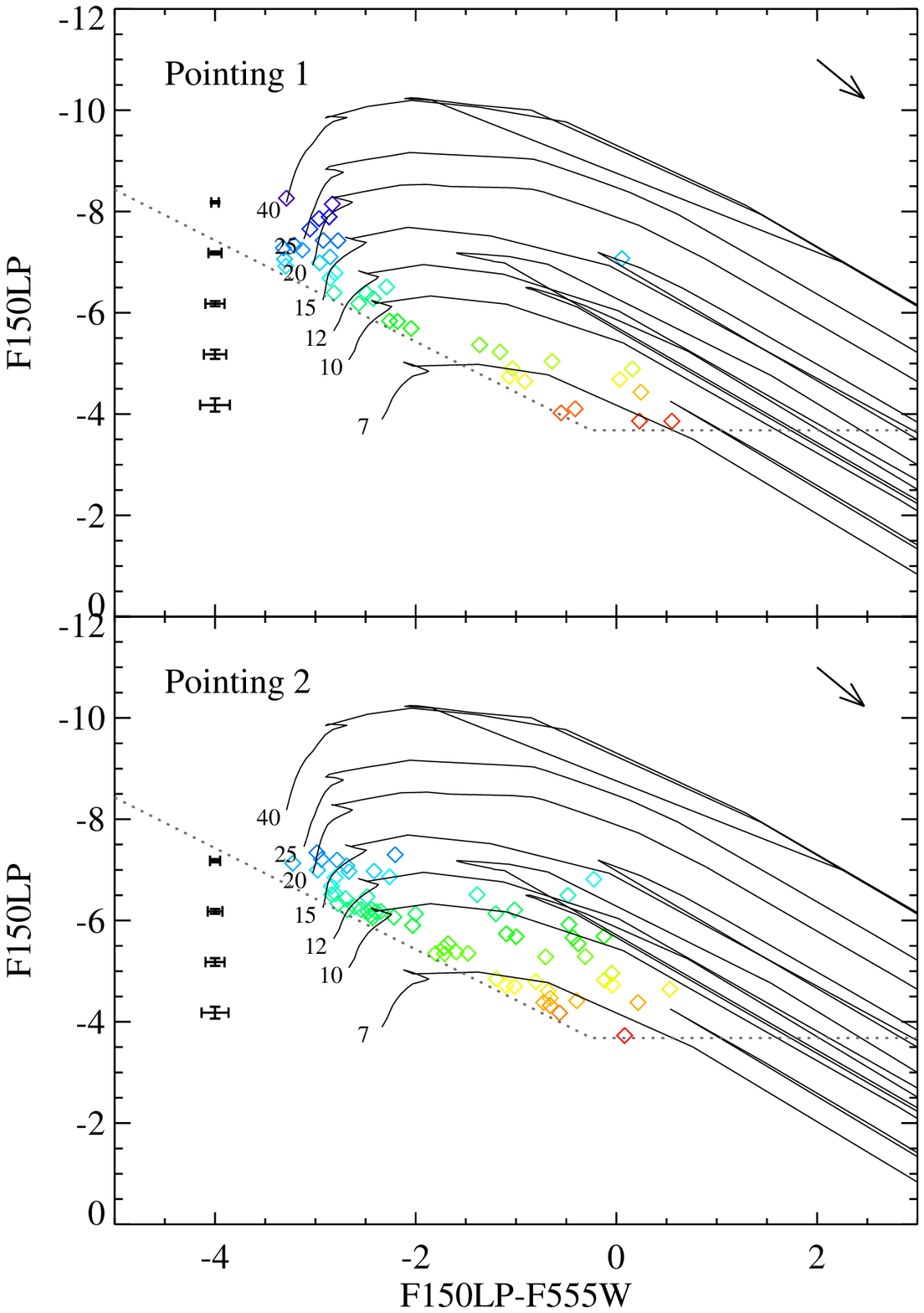}
\includegraphics[width=8cm]{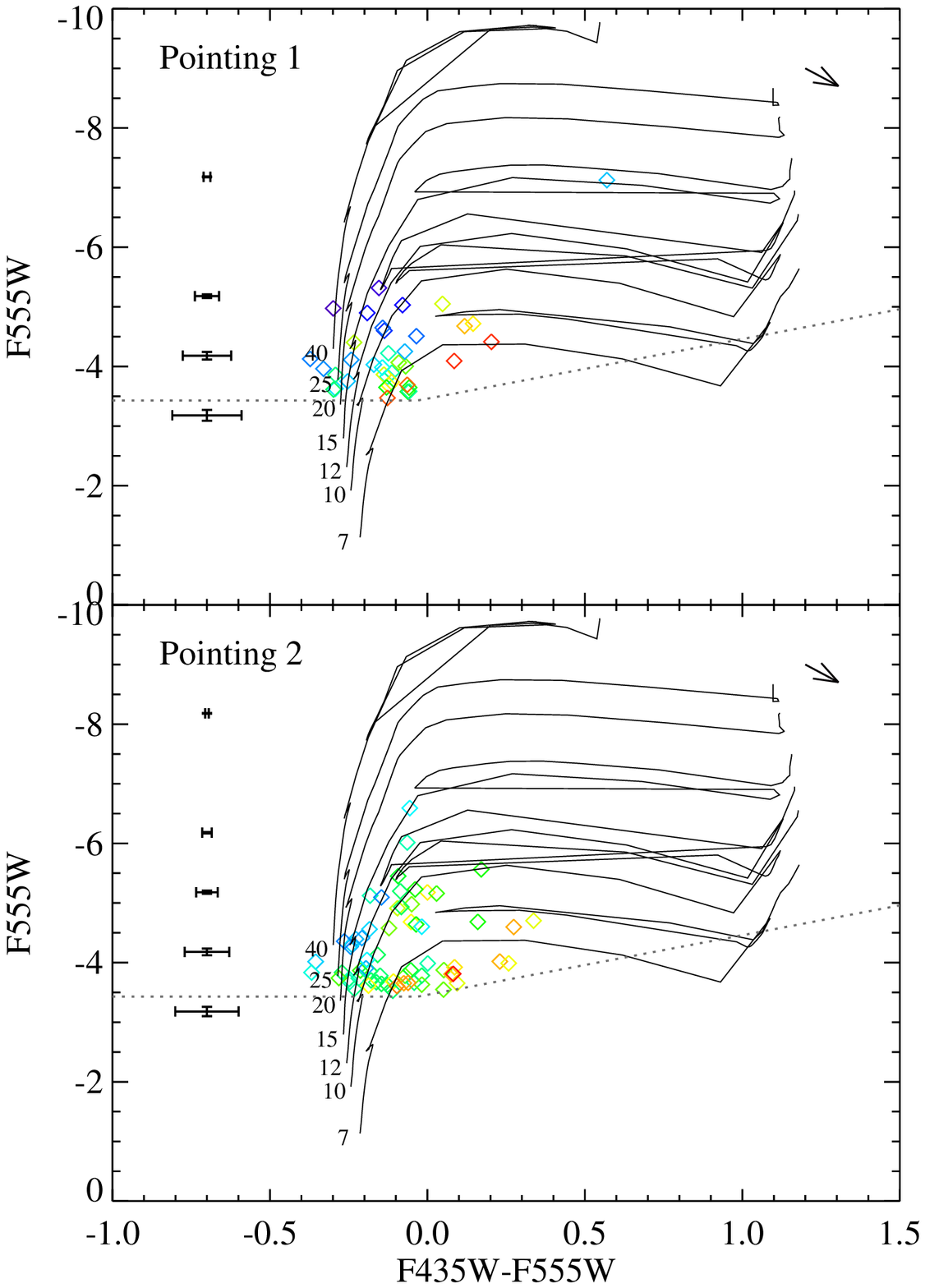}

\caption{Color-magnitude diagrams for the stellar FUV sources detected in the optical. Objects are color-coded by F150LP magnitude in both panels, with the scaling evident from the left panel. The dotted grey lines represent our completeness limits ($m_{\mathrm{F150LP}}=25.5$ and $m_{\mathrm{ F555W}}=m_{\mathrm{F435W}}=25.75$). An extinction vector for A$_{\mathrm{V}}=0.3$ is shown as are error bars appropriate for each magnitude bin. Stellar tracks from z=0.008 Geneva models are plotted with the stellar mass of each noted in small font. Pointing 1 demonstrates a more extended upper main sequence than Pointing 2, while Pointing 2 exhibits a higher proportion of sources evolving off the main sequence.}

\label{fig:colormag}
\end{center}
\end{figure*}

The stellar color-magnitude diagrams (CMDs; Fig.~\ref{fig:colormag}) qualitatively agree with the luminosity function derived SFHs, although they are limited in diagnostic power because only the brightest stellar FUV sources are optically detected (hence the far fewer number of stars plotted). The F150LP-F555W CMD (left panel) reveals that Pointing 1 contains more massive stars than Pointing 2, as expected based upon the star-forming spiral arm edge located in Pointing 1. Similarly, the recent ten-fold decrease of star-formation in Pointing 2 can explain the greater number of evolved sources found in its CMD compared to Pointing 1. We note that the F150LP-F555W CMD (left panel) is much better at distinguishing masses for young, massive stars which are extremely degenerate in the F435W-F555W CMD (right panel).

\subsection{Clusters}

\begin{figure}
\begin{center}
\includegraphics[width=7.5cm]{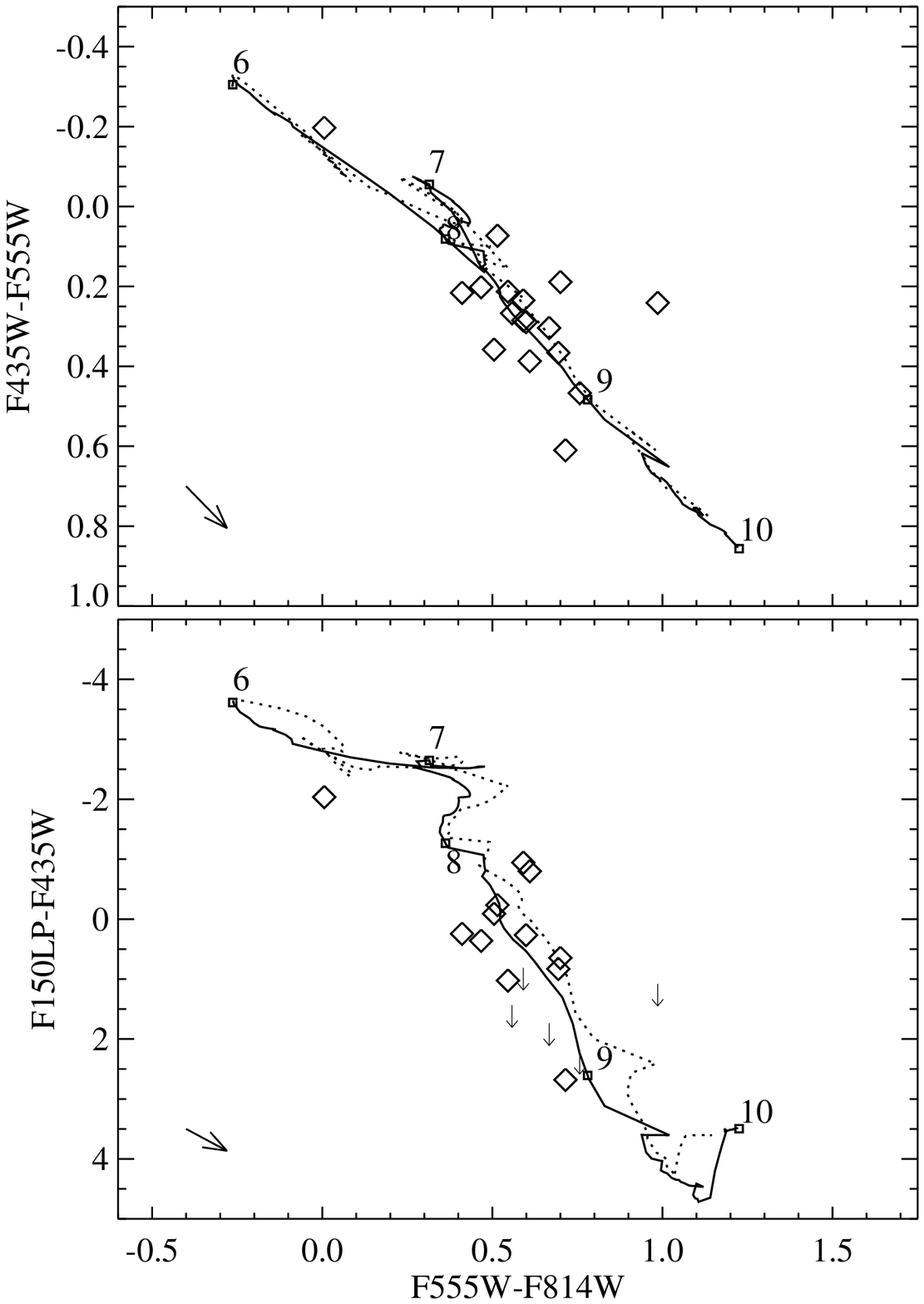}
\caption{Color-color diagrams for probable clusters in both fields. Lines show SSP evolutionary tracks at z=0.008 (solid) and 0.004 (dotted) from BC03. 
The arrow in the lower-left corner shows the direction an extinction of $A_{\mathrm{V}}$=0.3 would move data points. The single digits mark logarithmic ages along the cluster evolutionary track. }
\label{fig:clusters}
\end{center}
\end{figure}

We have two cluster samples based upon two different goals. The first is designed to help account for all the FUV light detected and is thus very inclusive. It consists of all non-stellar F150LP sources other than the two sources in each field specifically identified as  background galaxies. As a reminder, non-stellar F150LP sources were identified as those with optical counterparts with F555W FWHM above 2.35 pixels. Some of these sources may indeed be background galaxies and some may be very low signal-to-noise point sources with inaccurately measured FWHM.  The total amount of FUV light in these sources is 0.6 and 1.4\% in Pointing 1 and 2, respectively. Thus this category of sources does not appear to contribute significantly to the interarm FUV emission in M101.

The second sample is designed to investigate the properties of the clusters themselves. For this sample, we identify stellar clusters from the F435W and F555W images, since not all clusters are detected in the FUV. Cluster candidates are selected to have a FWHM larger than 2.5~pixels in both filters, and to be brighter than 25.5~mag in the F555W band. This selection yields seven probable clusters in Field 1 and ten in Field 2, eleven total of which are detected in the F150LP images. 

Fig.~\ref{fig:clusters} shows color-color plots comparing the measured colors of the clusters, denoted with diamonds or upper-limit arrows, with predictions from theoretical evolutionary models. A very small correction for foreground extinction was applied to the observed colors (approximately one tenth the length of shown extinction arrow). The solid (dotted) line represents the evolution of a Z=0.008 (Z=0.004) population as it ages \citep[][hereafter BC03]{bruzual03}, with the logarithmic age in years given above the line. The disk of M101 has a steep metallicity gradient \citep[e.g.,][]{li_y13}, and these metallicities should  approximate those at the location of our fields. The direction and length of the small arrow in the lower-left corner of each panel shows the expected change for an extinction of $A_V=0.3$, an upper limit to the expected extinction. The top panel of Figure~\ref{fig:clusters} plots optical colors ($F435W-F555W$ vs. $F555W-F814W$). While the data points are close to the model predictions, age and reddening are clearly degenerate in this color-color space.  The lower panel, which includes the FUV measurement, helps to break the age-extinction degeneracy. Here, extinction will move cluster colors off the model tracks. However, in this panel, the cluster colors are reassuringly close to the predicted ones, indicating that there is very little extinction towards these objects.

Estimating the ages of the clusters from Figure~\ref{fig:clusters}, we find that they primarily formed with a range of ages between 100~Myr to 1~Gyr ago. Only a single cluster has formed much more recently, within the past 10~Myr. The result that most of the clusters are of intermediate age is consistent with
their location between spiral arms, since most clusters are expected to form within and not between arms.
These clusters have likely moved away from their birth sites due to their rotation with the M101 disk.

We estimate the mass of each cluster from the F555W luminosity and the predicted age-dependent
mass-to-light ratio from the BC03 models.
The clusters have a range between $10^{3}$ up to a few times $10^{4}$ \Msun.
The fairly low masses and intermediate ages of
these clusters suggest that the outer M101 disk is a fairly hospitable place for clusters to survive. Galactic shear will be minimal due to the flat rotation curve at these radii  and tidal shocks from passing molecular clouds may be particularly infrequent. 
A full study of cluster masses and ages throughout M101 would be required
to more firmly establish the longevity of low-mass clusters in the outer disk.
It is also possible that some clusters are even lower metallicity than we have considered here (if they are low-metallicity ancient clusters). In this case, their ages would be older and masses larger than we have currently estimated. 

\section{Diffuse Light}
\subsection{Estimates of non-detected stars}

The emission from non-detected stars is the most uncertain portion of this accounting. Main-sequence stars emit in the FUV up to about 1~Gyr and the large number of these relatively lower-mass stars means they are a significant source of FUV light, despite their individual dimness. 

Our SFH fits only include information up to about 100~Myr (Section 3.1). We estimate emission from non-detected stars in three different ways.  The first way under-predicts the contribution from undetected stars by assuming that only stars younger than  the directly-probed 100~Myr contribute to the FUV flux. The second case 
 we consider our best estimate. It extrapolates the best fit SFR$_{\mathrm{final}}$ (or the inter-arm SFR) back to 1 Gyr, under the assumption that the high SFR$_\mathrm{initial}$ value is essentially a burst corresponding to passage through the spiral arm that does not last for the entire past 1 Gyr. (For Pointing 1 - NW which is best fit by a constant SFR, we assume a drop to one-tenth of the current SFR 100 Myr ago for this case.) The third case extrapolates this higher SFR$_\mathrm{initial}$ back to 1 Gyr and thus should be regarded as an upper limit to the amount of non-detected FUV emission from stars. These three assumptions result in the fractions graphically depicted for each region in Fig.~6 and tabulated for the best estimate (second method) in Table~2.

 As Fig.~6 shows, detected and non-detected stars can not account for all of the FUV emission in Pointing 1. The story is more ambiguous for Pointing 2. Pointing 2 is a fairly isolated region, far from a spiral arm. A constant low level of star formation might be expected for such a region, however, there is a clear sign of a significant decline (by a factor of 10) in approximately the last 10~Myr. Therefore, the best extrapolation of the SFR between 100~Myr and 1~Gyr is unclear, but the choices of SFR$_\mathrm{initial}$ and SFR$_\mathrm{final}$ likely bracket reality. In fact, using the higher SFR$_\mathrm{initial}$ can be ruled out, because it would require more FUV emission than is observed. 

\begin{table}[htdp]
\begin{center}
\caption{}
\begin{tabular}{lccc}
\hline
Region & P1 - NW & P1 - SE & P2\\
\hline
\flam(F150LP) & 1.36E-11 & 1.84E-11& 3.55E-11 \\
FUV diffuse SB  & 1.76E-5 & 1.57E-5 & 0.00\\
PAH SB  & 0.29 & 0.17 & 0.09\\
\hline
Fraction of FUV in: \\
Background galaxies  & 4.7E-5 & 6.2E-3 & 1.3E-3\\
Clusters  & 3.5E-5 & 4.2E-3 & 1.3E-2\\
Detected Stars  & 0.43 & 0.059 & 0.33 \\
Nondetected stars  & 0.27 & 0.32 & 0.52 \\
Diffuse  & 0.29 & 0.61 & 0.14\\

\hline
\end{tabular}
\end{center}

For $A_{\mathrm{V}}$ = 0.1 and SFR$_{\mathrm{final}}$ extrapolated between 100~Myr and 1~Gyr. Flux is in erg s$^{-1}$ cm$^{-2}$ \AA$^{-1}$. FUV surface brightness is in counts s$^{-1}$ pix$^{-1}$; PAH surface brightness is in MJy sr$^{-1}$.
\label{default}
\end{table}%

\begin{figure}
\begin{center}
\includegraphics[width=7.5cm]{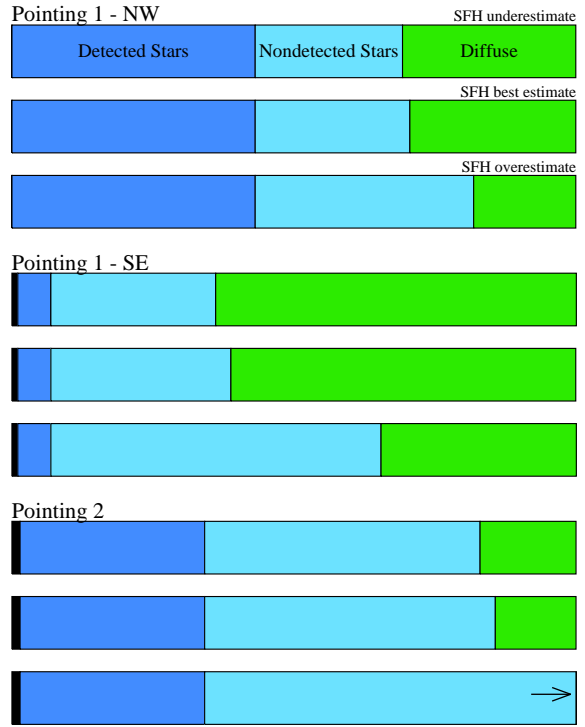}
\caption{Showing the effect of different SFH assumptions on non-detected stellar flux. For each region, the top bar shows the minimal correction of accounting only for non-detected stars from the past 100~Myr (underestimate). The next two bars show the fit SFR$_{\mathrm{final}}$ (best estimate) and SFR$_{\mathrm{initial}}$ (overestimate) for the past 100~Myr to 1~Gyr period. The small black region on the left-hand side depicts the fraction of FUV due to background galaxies and stellar clusters within the galaxy.}
\label{fig:bars}
\end{center}
\end{figure}

\subsection{Diffuse emission as scattered spiral arm light}

The radiative transfer models from \citet{witt00} predict scattered to stellar flux ratios of $F_{\mathrm{SCAT}}/F_{\star} = 0.1-0.6$ at an optical depth of $\tau_{\mathrm{V}}=0.5$ for clumpy Milky Way dust at 1614~\AA~ (6.2 $\mu$m$^{-1}$; see their Fig. 8). We note that the lower optical depth towards our regions ($\tau_{\mathrm{V}}\approx0.1$) should give slightly lower scattered to stellar fractions, but the perpendicular viewing direction should increase the same fraction due to increased probability of scattered light escape. Ratios of our measured diffuse to stellar flux ratios for our preferred star formation history are 0.4, 1.6 and 0.16 for Pointing 1 - NW, Pointing 1 - SE and Pointing 2. While Pointing 1 - NW and Pointing 2 are thus in the expected range, Pointing 1 - SE is higher. For this region, much of the scattered light probably does not originate within the delineated region (instead from the spiral arm included in Pointing 1 - NW) explaining such a high fraction of scattered to local stellar light.

We may also test for direct empirical evidence of scattered light in Pointing 1. As P1 - SE is immediately adjacent to a UV-bright spiral arm, the amount of light scattered should depend both on the local radiation density (and its direction) and on the density of scattering particles, here dust. 

We obtain dust surface densities from Aniano et al. (in prep.) who apply \citet{draine07a} models to the Spitzer and Herschel mid- to far-IR photometry of M101 (along with several other galaxies in the KINGFISH survey). A dust mass surface density of 1.2$\times10^{5}$ \Msun kpc$^{-2}$ is typical for Pointing 1.  Based on data for NGC~891 \citep{bianchi11}, NGC~4565 \citep{delooze12} and seven other edge-on spiral galaxies with Herschel data \citep{verstappen13}, we adopt a scale height of 200~pc resulting in a dust volume density of 3$\times10^{5}$ \Msun kpc$^{-3}$ in the disk midplane. Assuming uniformly distributed dust and a dust cross-section per gram from \citet{draine03a}, a FUV optical depth of $\tau=1$ is reached  within the midplane at only 210~pc, about a fifth of our SBC field-of-view. However, 200~pc above the midplane, $\tau=1$ occurs after FUV photons travel about 560~pc. Thus, if dust is uniformly distributed, it is difficult for FUV photons to traverse much more than about 500~pc, even at the lower dust volume densities of the outer disk. 

Given these considerations, we look for a higher fraction of scattered emission in the region adjacent to the spiral arm in Pointing 1. To test this, we break the SE region of Pointing 1 into an intermediate region within 210~pc of the spiral arm and a more distant region encompassing the remainder of the field. Unfortunately we cannot refit for the SFH separately in these two regions, because so few stars are detected. Instead we assume the same correction factor for non-detected stellar flux in both. Performing this calculation, we find that the region adjacent to the spiral arm actually has less fractional diffuse emission and a lower diffuse FUV surface brightness. This is the opposite of the expected trend if scattered light dominates the diffuse FUV emission observed in our field. 

While this simple test fails to establish a clear signature of scattered light, if the stellar population of the outer region is older than that of the inner region, the diffuse fraction in the above analysis would be underestimated for the inner region and overestimated for the outer region, thus still leaving room for the expected pattern from scattering.

\subsection{Other sources of diffuse FUV}

Besides locally scattered FUV from M101, other potential sources for the diffuse FUV emission include:  foreground (Galactic) emission, background (extragalactic) emission, or emission from the warm ionized medium (WIM) within M101. 

Measurements of the full-sky diffuse FUV reveal that it correlates well with \hi column density and 100 \micron dust emission, implying a foreground origin \citep[e.g.,][]{paresce80a, jakobsen97}.   Based upon UV measurements from the Spectroscopy of Plasma Evolution from Astrophysical Radiation (SPEAR) instrument, \citet{seon11} document the relation between atomic hydrogen column density and $I_{\mathrm{FUV}}$ in their Fig.~21. Given the foreground extinction predicted by the \citet{schlafly11} model, we estimate that the foreground \hi column density is $4.3\times10^{19}$ cm$^{-2}$ in the direction of M101. At this low \hi column density, the $I_{\mathrm{FUV}}$ curve gives a value of 350$^{+200}_{-100}$ photons s$^{-1}$ cm$^{-2}$ sr$^{-1}$ \AA$^{-1}$. 
Background galaxy emission is either in the form of identifiable background galaxies apparent in gaps of the GMC cloud cover (as the two galaxies in each field we have identified in the FUV) or is unlikely to have passed through the M101 disk \citep{holwerda07a,holwerda07b}. Hence, we conservatively take 550 photons s$^{-1}$ cm$^{-2}$ sr$^{-1}$ \AA$^{-1}$ as an upper limit to the foreground and consider background FUV emission negligible. This value equates to  $6.8\times10^{-8}$ counts s$^{-1}$ pix$^{-1}$ in our maps, approximately 500 times below the FUV surface brightnesses that we measure.

Within the Galaxy, the WIM contribution to the diffuse FUV from hydrogen two-photon emission is estimated to be about  4-9\% \citep{seon11}. A proportionality with \halpha emission was first put forth by \citet{reynolds92}, here we use $I_{2\mathrm{\gamma}} = 57.4\;I_{\mathrm{H\alpha}}$, where $I_{2\mathrm{\gamma}}$ is in photons s$^{-1}$ cm$^{-2}$ sr$^{-1}$ \AA$^{-1}$, $I_{\mathrm{H\alpha}}$ is in Rayleigh and the electron temperature assumed for the WIM is 8000K. Using an \halpha image of M101 obtained with the Kitt Peak National Observatory \citep{thilker02}, we determine the average \halpha surface brightness to be 2.9 and 1.6 Rayleigh, for Pointing 1 and 2, respectively. Thus the expected contribution from two-photon WIM emission is $2.0\times10^{-8}$ and $1.1\times10^{-8}$ counts s$^{-1}$ pix$^{-1}$, respectively, also negligible. 

Recombination emission from Ly$\alpha$ can pump molecular hydrogen which then emits in the FUV band. This phenomenon is observed in planetary nebulae as well as young star forming regions. While the pumped-H$_{2}$ contribution  to the diffuse FUV is around 10\% in the Taurus-Perseus-Auriga complex \citep{lim13}, we don't expect such a large contribution in our regions due to the lack of molecular hydrogen and the dimness of the observed \halpha recombination emission. 

Having ruled out these other potential sources of diffuse FUV emission, we conclude that remainder of interarm FUV emission after stellar sources are removed is due to locally scattered light within M101. 

\section{Conclusions}

Using deep and well-resolved FUV images from HST, we constrain the young stellar populations in two interarm pointings in the outskirts of M101's disk. Hundreds of point sources are detected in both F150LP images, allowing us to create luminosity functions. These FUV luminosity functions  have different shapes in the three regions studied. These shapes determine the best-fit recent star-formation history. In the two truly interarm regions, we find evidence of truncated star-formation histories, indicating very little if any star-formation is presently occurring in these regions. 

%As the majority of these stars are completely invisible in the optical, this approach is really a new way to constrain recent SFHs within such small regions.

Neither background galaxies nor stellar clusters significantly contribute to the FUV light in interarm regions, amounting to a combined $<1.5\%$. However, several low mass ($<10^{4}$ \Msun) clusters are detected in each pointing. Almost all have ages between 100~Myr - 1~Gyr, based on a comparison with stellar population tracks. The detection of such clusters signals that the disruption of low-mass clusters is not extremely efficient in the outer disk of M101.

In both interarm regions studied, significant amounts of FUV light remain after the detected stellar sources, clusters, and background galaxies are removed. The flux remaining is consistent with predictions of scattered light from radiative transfer models. In Pointing 1, which is immediately adjacent to a spiral arm, we find that scattered light from the arm must contribute significantly to the total FUV in that field ($\approx 60$\%). Meanwhile, Pointing 2, which is further from any spiral arm or bright star-forming knots, has an upper bound of 16\% scattered light. Thus the FUV emission in interarm regions in M101 is partly due to individual FUV-bright stars and partly due to light scattered from the spiral arms. The latter contribution is only significant immediately adjacent to spiral arms.

\acknowledgements{The authors would like to thanks the referee for thoughtful comments which improved the manuscript. This paper is based on observations taken with the NASA/ESA Hubble Space Telescope obtained at the Space Telescope Science Institute, which is operated by AURA, Inc., under NASA contract NAS5-26555. These observations are associated with the program numbers 9490 and 11147.}

\bibliographystyle{apj}
\bibliography{master-apj}

\clearpage

\appendix

\section{SBC Dark Current Removal}

In the ACS Data Handbook, the SBC dark current is listed as $0.8-1\times10^{-5}$  counts s$^{-1}$ pixel$^{-1}$. However, it also mentions that the dark current increases by about a factor of 5 over 2 hours (Section 4.4.1), as the detector temperature increases. SBC data obtained of the Hubble Ultra Deep Field also indicates this temperature-related increase in dark current, further noting that the dark current exhibits structure with a peak near the detector center \citep{teplitz06}. 

\begin{figure*}
\begin{center}
\includegraphics[width=15cm]{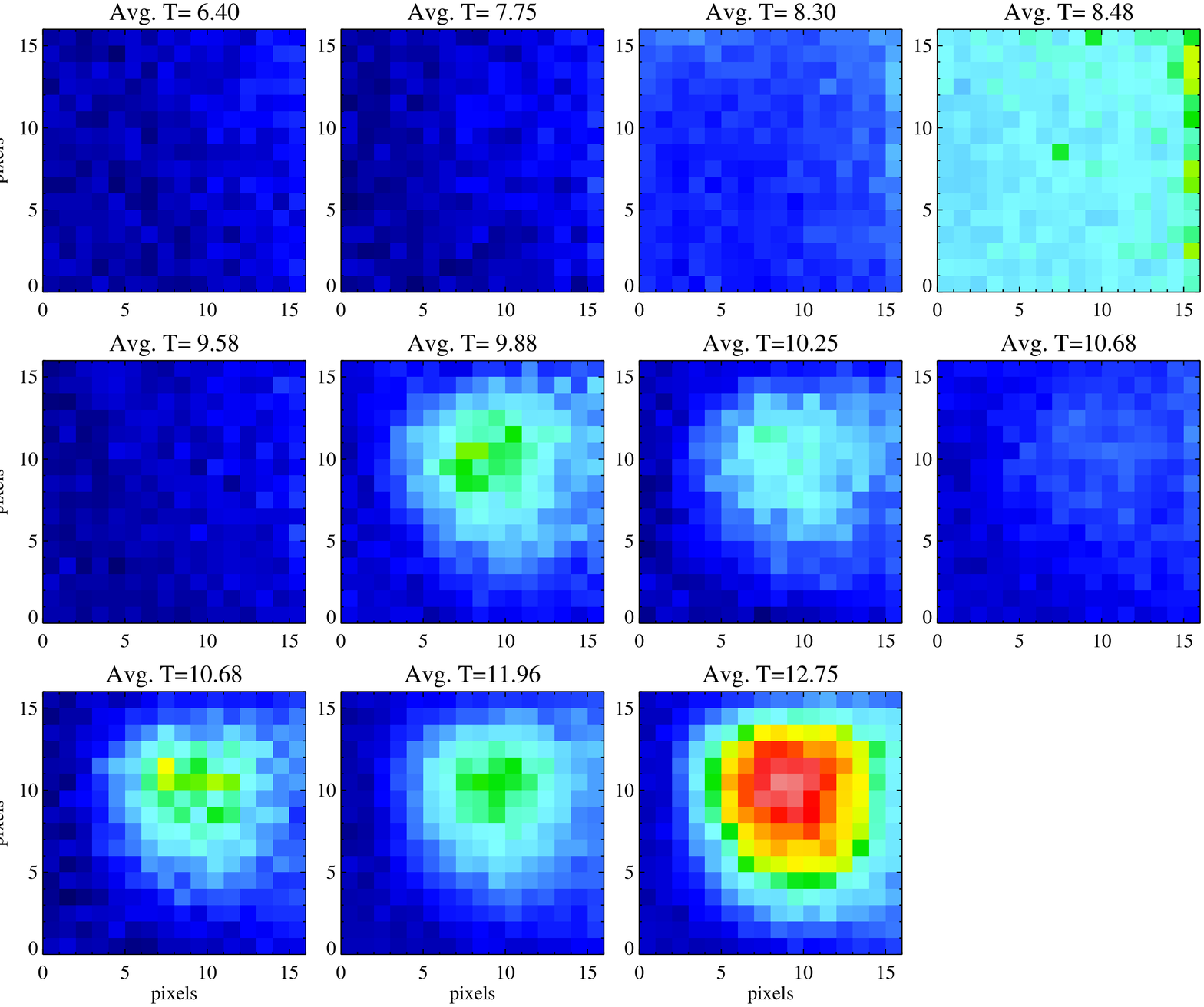}
\caption{Dark current calibration images from 2008 organized by average detector temperature from left to right and top to bottom. The dark current shows a clear structure at higher temperatures, but the strength of the feature is not monotonic with average temperature. At lower temperatures variations are also present, most noticeably the globally higher value of the Avg. T = 8.48 observation. }
\label{fig:darks}
\end{center}
\end{figure*}

We find that the dark current can be modeled by two components based on all the darks obtained with the SBC for 2008 (the year of our observations). These are shown in Fig.~\ref{fig:darks}, having been binned to a 16 by 16 grid in order to increase the signal of the dark current. In this figure, the high-temperature dark images show a peaked structure that is absent in images taken when the detector temperature lower than about 9.5$^{\circ}$C. Unfortunately, the detector temperature (as measured by the average of the two MDECODT keywords) does not appear to be the only parameter governing the dark current rate. Even at low temperatures, the dark current level can vary by a factor of two, although the structure of the dark-current in the low temperature darks appears similar, increasing from left to right. Based upon this inspection, we assume the dark current has a stable low-temperature structure (but arbitrary scaling) and an additional high-temperature component (with arbitrary structure and scaling).

\begin{figure}[htbp]
\begin{center}
\includegraphics[width=7.5cm]{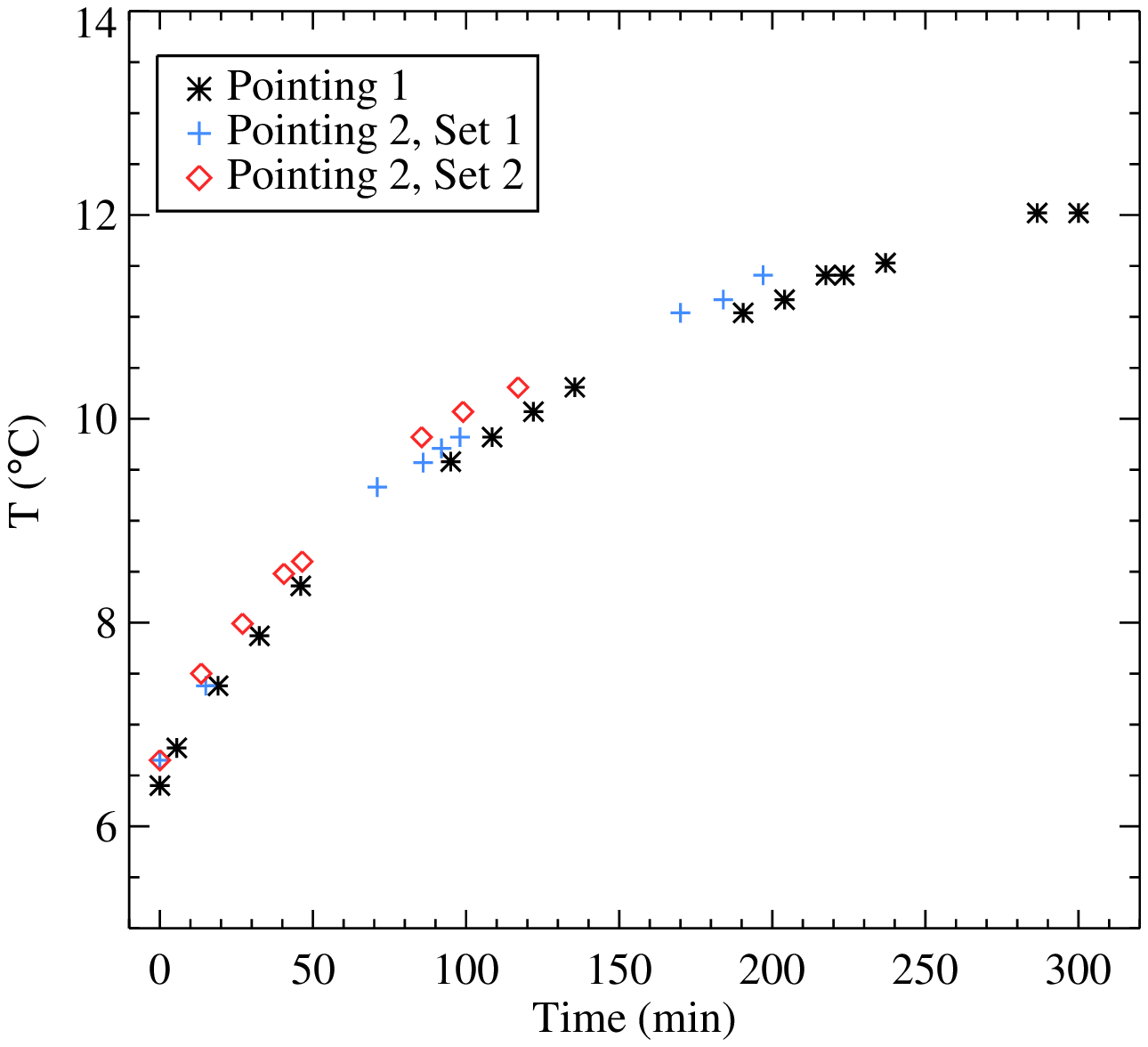}
\caption{Detector temperature as a function of time since the SBC was turned on for our observations. Each exposure set has a similar profile with time and shows clear gaps where M101 was presumably unobservable. Sub-exposure sets are grouped using these breaks to determine maps of the hot dark current.}
\label{fig:ourtemp}
\end{center}
\end{figure}

First we obtain high-temperature dark current maps by comparing our data taken at high temperatures (above 9$^{\circ}$C) to those at low temperatures.  Fig.~\ref{fig:ourtemp} shows the detector temperatures of our sub-exposures as a function of time since the detector was turned on. The gaps show times when M101 was unobservable during the orbit, so we average the two to five sub-exposures within these naturally-divided time blocks in order to increase the signal-to-noise. As the first time block of exposures is always quite cold (under 8$^{\circ}$C), we assume these averaged frames have no high-temperature component dark current. For the higher-temperature averaged images (three such blocks for Pointing 1, two for Pointing 2 - Set 1 and only one for Pointing 1 - Set 2), we directly subtract the cold exposure in order to obtain a map of the high-temperature dark current (always working on the coarse 16 x 16 grid, so slight positional uncertainties are not important). We then subtract these high-temperature dark current images from each sub-exposure within each corresponding high-temperature time block. All the frames for an observing day are then combined, producing an image with only the cold dark current component still present. 

We determine the scale of the cold dark current by comparing our SBC observations with dark-current free GALEX data (note the cold dark current structure is assumed to have the form of the averaged cold dark calibration frames). The SBC F150LP band is slightly redder than the GALEX FUV band (pivot wavelength of 1516\AA\xspace compared to 1612\AA\xspace). Because of this difference, we compute a color correction, based upon stars from the Bruzual \& Charlot 1995 library, as provided by the STSDAS SYNPHOT package. We select stars that match the FUV-NUV colors we observe in the two fields, which range from spectral type A2 to B7. We use E(B-V)=0.0 and 0.5 and a Milky Way extinction curve, expecting these E(B-V) values to bracket the expected extinction in these outer regions of M101. The computed color correction is parametrized as:

\begin{equation}
\frac{F_{\lambda}(SBC)}{F_{\lambda}(FUV)}=1.66-0.594\frac{F_{\lambda}(FUV)}{F_{\lambda}(NUV)}+0.129\left(\frac{F_{\lambda}(FUV)}{F_{\lambda}(NUV)}\right)^{2}.
\end{equation}
$F_{\lambda}$ values are computed from provided PHOTFLAM values from \citet{morrissey07} and the ACS Zeropoints webpage. Typical color correction factors range from 1.03 to 1.10, with the maximum of 1.23. 

\begin{figure*}
\begin{center}
\includegraphics[width=5.5cm]{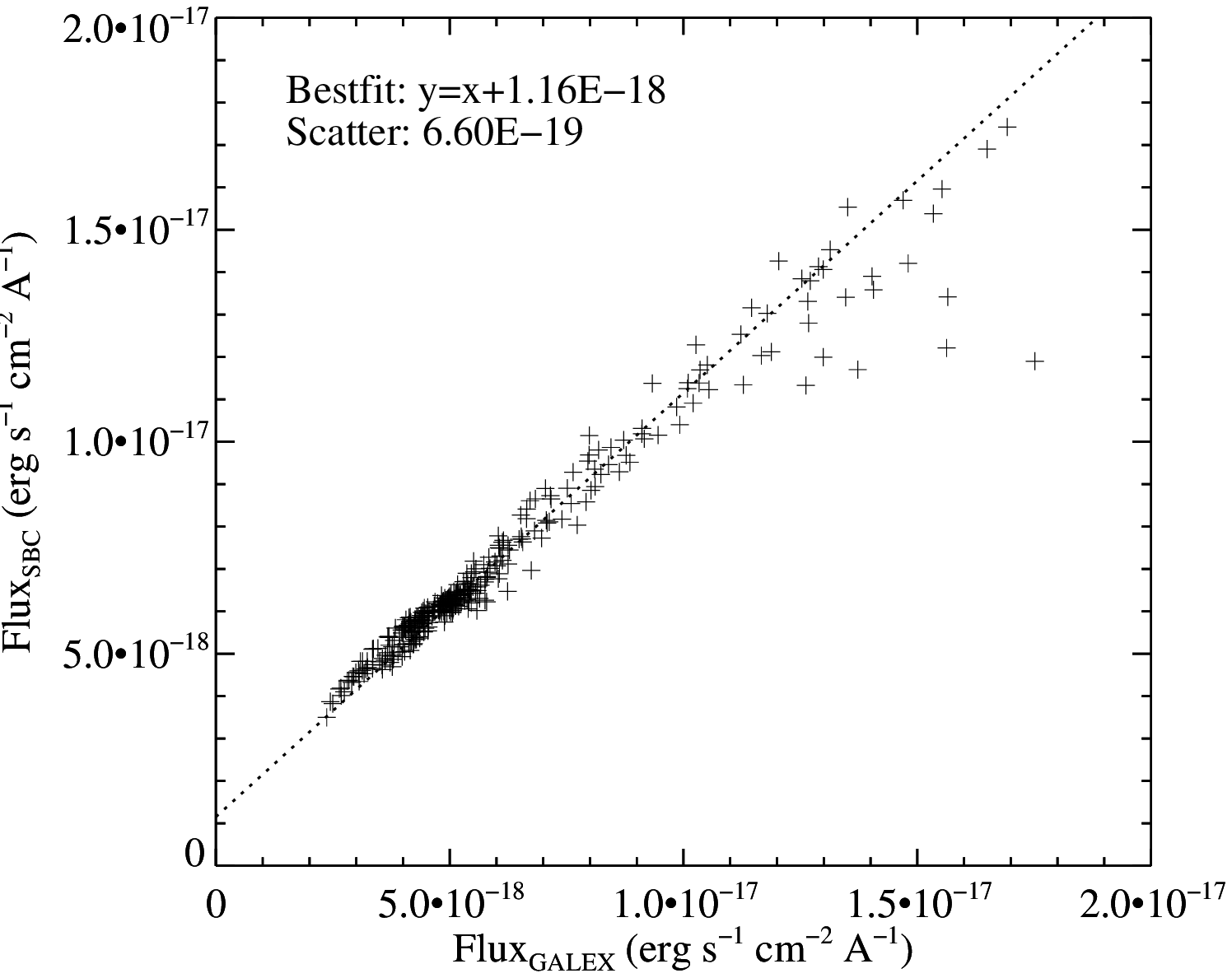}
\includegraphics[width=5.5cm]{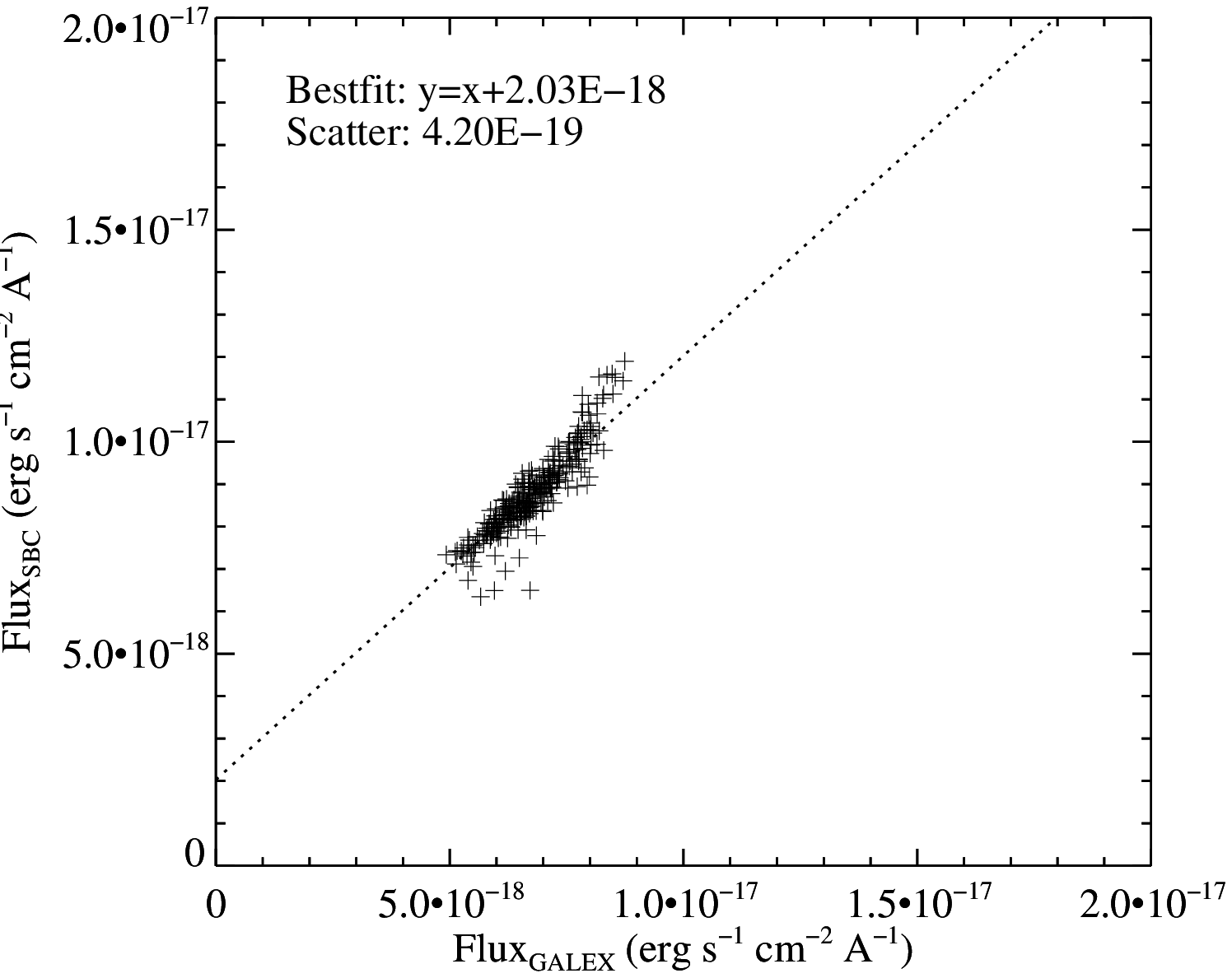}
\includegraphics[width=5.5cm]{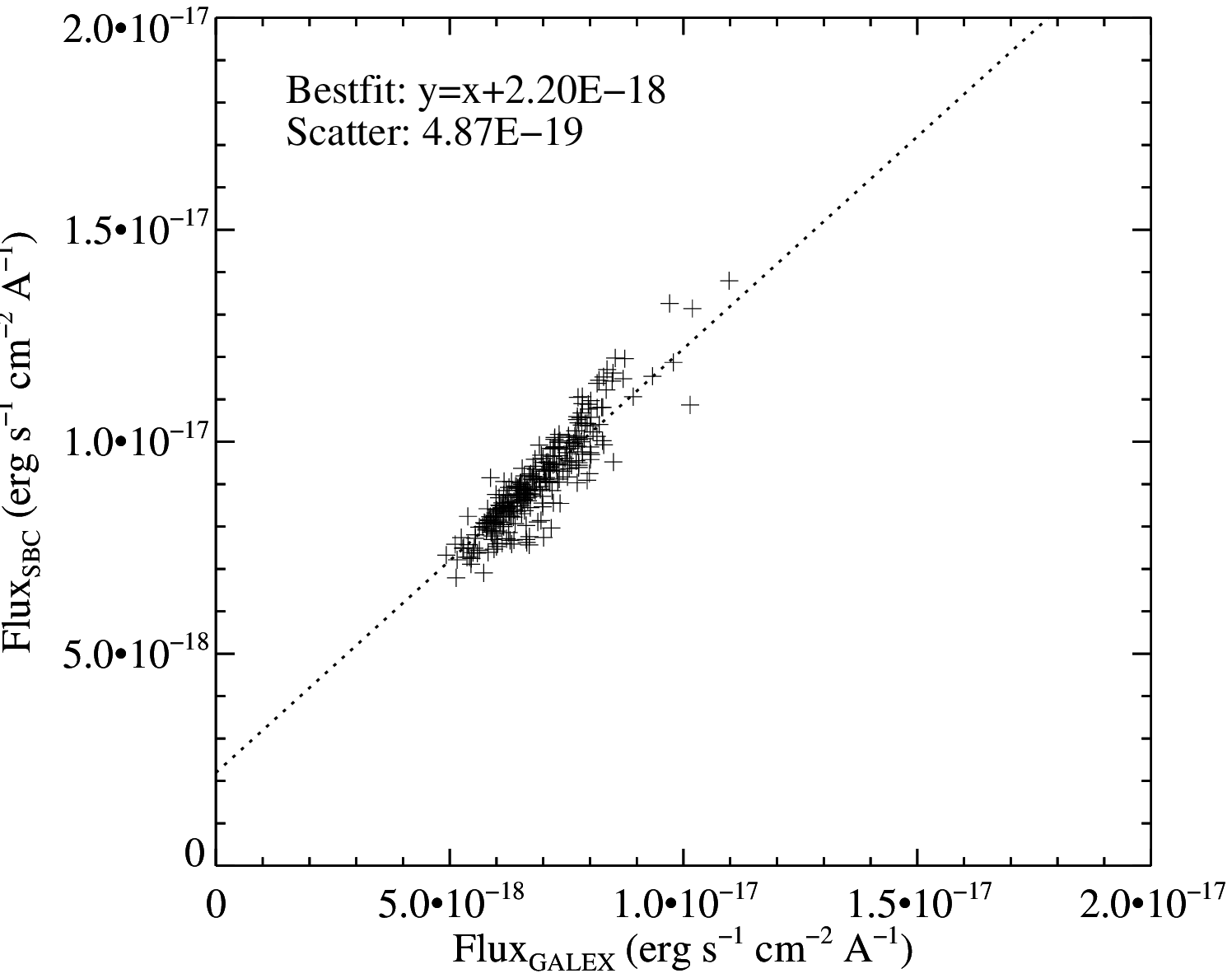}
\caption{For the three different observing sets, the SBC measured flux density per 1.5\arcsec pixel compared to the predicted flux density from GALEX (using the FUV-NUV color correction to the FUV flux). Left to right: Pointing 1, Pointing 2 - Set 1 and Pointing 2 - Set 2. }
\label{fig:sbc_vs_galex}
\end{center}
\end{figure*}

Applying this color correction  to the GALEX FUV images leaves us with predicted dark-current free images for SBC F150LP at the GALEX resolution.  Convolving the three SBC images with the GALEX FUV PSF using both the PSF and the IDL program of \citet{aniano11} and then resampling to the 1.5\arcsec\xspace pixels of the GALEX image, we may directly compare the SBC images to the GALEX prediction. 
To determine the dark current level, we fit a one-to-one line to the pixel flux density of the observed versus predicted image as seen in Fig.~\ref{fig:sbc_vs_galex}. Such a one-to-one line is generally a good fit (some bright points off the relation are probably due to insufficient color correction for areas with bright OB stars), thus the offset provides the average cold dark current value. These offsets are $1.16\pm0.04\times10^{-18}$, $2.03\pm0.02\times10^{-18}$ and $2.20\pm0.03\times10^{-18}$ erg s$^{-1}$ cm$^{-2}$ \AA$^{-1}$ in our three SBC exposure sets. These correspond to $0.74\times10^{-5}$, $1.29\times10^{-5}$ and $1.40\times10^{-5}$ counts s$^{-1}$ in the detector pixels, comparable or slightly higher than the values given in the ACS handbook. We apply this scale to a normalized master cold dark computed from all of the low-temperature 2008 darks in order to remove the cold dark current from each of the sub-exposures. This dark subtraction accounts for approximately a quarter of the original flux in our images.

\label{lastpage}

\end{document}